\documentclass[aps,preprint,amsmath,amssymb,superscriptaddress,nofootinbib]{revtex4}
\usepackage{multirow}
\usepackage{graphicx}
\usepackage{mathtools}
\usepackage[makeroom] {cancel}

\begin{document}

\title{Projected Sensivity to Dimension-6 Triple Gauge Couplings at the FCC-hh}

\author{V. Ari}
\email[]{vari@science.ankara.edu.tr} \affiliation{Department of Physics, Ankara University, 06100, Ankara, Turkey}

\author{V. Cetinkaya}
\email[]{volkan.cetinkaya@dpu.edu.tr} \affiliation{Department of Physics, K\"{u}tahya Dumlup{\i}nar University, 43100, K\"{u}tahya, Turkey}

\author{M. K\"{o}ksal}
\email[]{mkoksal@cumhuriyet.edu.tr} \affiliation{Department of Physics, Sivas Cumhuriyet University, 58140, Sivas, Turkey}

\author{O. Cakir}
\email[]{ocakir@science.ankara.edu.tr} \affiliation{Department of Physics, Ankara University, 06100 Ankara, Turkey}

\begin{abstract}

In this study, we investigate the process $pp \rightarrow W^{\pm} \gamma$ for the physics potential of the FCC-hh with $\sqrt{s}=100$ TeV to
examine the anomalous $WW\gamma$ couplings defined by three CP-conserving and two CP-violating effective
operators of dimension-6. The analysis containing the
realistic detector effects is carried out in the mode where $W^{\pm}$ bosons in the final state decay into the leptonic channel.
The best sensitivities obtained from the process $pp \rightarrow W^{\pm} \gamma$ on the anomalous
couplings $C_{WWW}/ \Lambda^{2}$ and $C_{W,B}/ \Lambda^{2}$ determined by CP-conserving effective Lagrangians
are $[-0.01; 0.01]$ TeV$^{-2}$ and $[-0.88; 0.88]$ TeV$^{-2}$,
while $C_{\tilde{W}WW}/ \Lambda^{2}$ and $C_{\tilde{W}}/ \Lambda^{2}$ couplings defined by
CP-violating effective Lagrangian are obtained as $[-0.03; 0.03]$ TeV$^{-2}$ and $[-0.47; 0.47]$ TeV$^{-2}$ at the FCC-hh with $\sqrt{s}=100$ TeV, $L_{int}=30$ ab$^{-1}$. However, if the systematic uncertainty is included, we obtain reduced sensitivities on the anomalous $WW\gamma$ coupling . The results are compared for assumed systematics of $5\%$ and $10\%$.
\end{abstract}

\maketitle

\section{Introduction}

Gauge boson self-interactions are exactly described by the non-abelian gauge symmetry of the Standard Model (SM). These interactions are important for understanding the gauge structure of the SM and new physics studies beyond the SM. Any deviations from the SM predictions on gauge boson self-interactions such as triple gauge boson couplings can give us important information related to the existence of new physics beyond the SM. To probe possible deviations on the triple gauge boson couplings arising from new physics in a model-independent way using the effective Lagrangian method is ordinarily a usual way. The effective Lagrangian including the higher operators than dimension-4 is given as follows

\begin{eqnarray}
\mathcal{L}_{eff}&=\mathcal{L}_{SM}+\Sigma_{D> 4}\Sigma_{i}\frac{C_{i}^{D}}{\Lambda^{D-4}}\mathcal{O}_{i}^{(D)}.
\end{eqnarray}
Here, $C_{i}$ shows the coefficients of the dimension-D operator, $\Lambda$ is the new physics scale, $\mathcal{O}_{i}$ is the desired operators. Avoiding the lepton and baryon number violation the effective Lagrangian becomes \cite{ef}

\begin{eqnarray}
\mathcal{L}_{eff}&=\mathcal{L}_{SM}+\Sigma_{i}\frac{C_{i}^{6}}{\Lambda^{2}}\mathcal{O}_{i}^{(6)}+\Sigma_{i}\frac{C_{i}^{8}}{\Lambda^{4}}\mathcal{O}_{i}^{(8)}+...
\end{eqnarray}
which includes only the even order effective operators. It is easily seen from the above equation that if we want to examine the effects of dimension-D operators (D = 6,8, ...) at low energies, it gradually decreases.

As known, the kinetic terms of the gauge fields in the SM produce the interactions of triple $WWZ$ and $WW\gamma$ gauge bosons at the tree-level due to the non-abelian nature of $SU(2)_{L}$ symmetry. However, there are no neutral triple $ZZZ$, $ZZ\gamma$, $Z\gamma\gamma$ and $\gamma\gamma\gamma$ gauge couplings at the tree-level in the SM. On the other hand, as mentioned above, the effective higher dimension operators may induce and modify the triple gauge boson interactions in the SM. To the lowest order (dimension-6), the operators contributing to $WWZ$ and $WW\gamma$ couplings, respecting the SM gauge symmetry, are \cite{1,2,3}

\begin{eqnarray}
{\cal O}_{WWW}=\text{Tr}\left[W_{\mu\nu}W^{\nu\rho}W_\rho^\mu\right]\,,
\end{eqnarray}
\begin{eqnarray}
{\cal O}_{W}=\left(D_\mu\Phi\right)^\dagger W^{\mu\nu}\left(D_\nu\Phi\right)\,,
\end{eqnarray}
\begin{eqnarray}
{\cal O}_{B}=\left(D_\mu\Phi\right)^\dagger B^{\mu\nu}\left(D_\nu\Phi\right)\,,
\end{eqnarray}
\begin{eqnarray}
{\cal O}_{\tilde{W}WW}=\text{Tr}\left[ \tilde{W}_{\mu\nu}W^{\nu\rho}W_\rho^\mu\right]\,,
\end{eqnarray}
\begin{eqnarray}
{\cal O}_{\tilde{W}}=\left(D_\mu\Phi\right)^\dagger \tilde{W}^{\mu\nu}\left(D_\nu\Phi\right)\,,
\end{eqnarray}

where $\Phi$ represents the Higgs doublet field. $D_\mu$, $W_{\mu\nu}$ and $B_{\mu\nu}$ are defined as

\begin{eqnarray}
D_\mu \equiv \partial_\mu\,+\,i\frac{g^\prime}{2}B_\mu\,+\,i g W_\mu^i\frac{\tau^i}{2}\, ,
\end{eqnarray}

\begin{eqnarray}
W_{\mu\nu}=\frac{i}{2}g\tau^i\left(\partial_\mu W_\nu^i - \partial_\nu W_\mu^i + g\epsilon_{ijk}W_\mu^j W_\nu^k \right)\,,
\end{eqnarray}

and

\begin{eqnarray}
B_{\mu\nu}=\frac{i}{2}g^\prime\left(\partial_\mu B_\nu - \partial_\nu B_\mu\right)\,.
\end{eqnarray}

Here, $\tau^i$ shows the $SU(2)_I$ generators with Tr $[\tau^i \tau^j]=2 \delta^{ij}$ $\left(i,j=1,2,3 \right)$. The $g$ and $g^\prime$ are $SU(2)_I$ and $U(1)_Y$ couplings, respectively.

The operators in Eqs. (3-7) are invariant under $SU(2)_I\otimes U(1)_Y$ gauge symmetry. To establish SM couplings or to capture new physics beyond the SM regardless of any symmetry, one has to go beyond the SM gauge symmetry. Therefore, one can simply think the Lorentz invariance and $U(1)$ symmetry to establish more general form factors as a function of the momentum concerned in a given vertex. In the form factor formalism, the effective Lagrangian for $WW\gamma$ interaction can be given by \cite{4}

\begin{eqnarray}
\label{eq.11}
{\cal L}_{WW\gamma}&=&ig_{WW\gamma}\Big[{g_1^{\gamma}}\left({W_{\mu\nu}^{+}}{W_{\mu}^{-}}A_{\nu}-{W_{\mu\nu}^{-}}{W_{\mu}^{+}}A_{\nu}\right) \nonumber \\
&+&\kappa_{\gamma}{W_{\mu}^{+}}{W_{\nu}^{-}}A_{\mu\nu}+\frac{\lambda_{\gamma}}{M_W^2}{W_{\mu\nu}^{+}}{W_{\nu\rho}^{-}}A_{\rho\mu} \nonumber \\
&+&ig_4^\gamma {W_{\mu}^{+}}{W_{\nu}^{-}}\left( \partial_\mu A_\nu+\partial_\nu A_\mu\right) \\
&-&ig_5^\gamma \epsilon_{\mu\nu\rho\sigma}\left({W_{\mu}^{+}}\partial_\rho{W_{\nu}^{-}}-\partial_\rho{W_{\mu}^{+}}{W_{\nu}^{-}}\right)A_\sigma \nonumber \\
&+&\tilde{\kappa}_{\gamma}{W_{\mu}^{+}}{W_{\nu}^{-}}\tilde{A}_{\mu\nu}+\frac{\tilde{\lambda}_{\gamma}}{M_W^2}{W_{\lambda\mu}^{+}}{W_{\mu\nu}^{-}}\tilde{A}_{\nu\lambda} \Big]\,, \nonumber
\end{eqnarray}
where $g_{WW\gamma}=-e$ and $\tilde{A}_{\mu\nu}=\frac{1}{2}\epsilon_{\mu\nu\rho\sigma}A^{\rho\sigma}$. Here, $A^{\mu\nu}=\partial^\mu A^\nu - \partial^\nu A^\mu$ is the field strength tensor for photon.  In above equation, $g_1^{\gamma}$, $\kappa_{\gamma}$ and $\lambda_{\gamma}$ anomalous parameters are both C and P conserving while $g_4^\gamma$, $g_5^\gamma$, $\tilde{\kappa}_{\gamma}$ and $\tilde{\lambda}_{\gamma}$ anomalous parameters are C and/or P violating. The anomalous couplings are described by $\Delta \kappa_{\gamma}=\lambda_{\gamma}=0$ ($\Delta \kappa_{\gamma}=\kappa_{\gamma}-1$) at the tree-level in the SM. Nevertheless, CP-violating interactions can be confined separately to specially designed CP-odd observables that are insensitive to CP-even effects. For this reason, the CP-violating and CP-conserving interactions can be distinguished from each other. In this case, $\kappa_{\gamma}$ and $\lambda_{\gamma}$ couplings can be transformed into $c_{WWW}/\Lambda^2$, $c_{W}/\Lambda^2$ and $c_{B}/\Lambda^2$ couplings as follows \cite{3}

\begin{eqnarray}
\label{eq.12}
{\kappa_\gamma}=1+\left(c_W+c_B\right)\frac{m_W^2}{2\Lambda^2}\,,
\end{eqnarray}
\begin{eqnarray}
\label{eq.13}
{\lambda_\gamma}=c_{WWW}\frac{3g^2m_W^2}{2\Lambda^2}\,,
\end{eqnarray}
\begin{eqnarray}
\label{eq.14}
{\tilde{\kappa}_\gamma}=c_{\tilde{W}} \frac{m_W^2}{2\Lambda^2}\,,
\end{eqnarray}
\begin{eqnarray}
\label{eq.15}
{\tilde{\lambda}_\gamma}=c_{\tilde{W}W W}\frac{3g^2m_W^2}{2\Lambda^2}\,.
\end{eqnarray}

Here, $c_{WWW}/\Lambda^2$, $c_{W}/\Lambda^2$ and $c_{B}/\Lambda^2$ couplings represent the possible deviations for $WW\gamma$ vertex in the SM. These parameters are equal to zero in the SM.

Many studies in the literature that theoretically include the anomalous $WW\gamma$ and $WWZ$ couplings \cite{ete,ete1,ete2,ete3,6,7,8,9,10,11,12,13,14,15,16,17,18,19,20,21,22,23,24,25,26,27,28,29,30}. The anomalous $WW\gamma$ couplings have been also studied experimentally at the LEP \cite{31,32,33}, the Tevatron \cite{34,35,36,37} and the LHC \cite{38,39,40,cms}. In Table~\ref{tab1}, the best limits are given at 95$\%$ C.L. on $c_{\tilde{W}WW}/\Lambda^2$, $c_{WWW}/\Lambda^2$, $c_{\tilde{W}}/\Lambda^2$, $c_{W}/\Lambda^2$ and $c_{B}/\Lambda^2$ parameters obtained from the experiments.

After the completion of the LHC and high center-of-mass energy and luminosity LHC physics programs, the future circular collider projects that are the FCC-ee, the FCC-eh and the FCC-hh are proposed to precisely measure the electroweak symmetry breaking mechanism of the SM and new physics effects beyond the SM. Triple Gauge Couplings (TGCs) have been studied \cite{li,billur,bian} at future colliders. Among these colliders, the FCC-hh has an energy scale that increases by about 7 times depending on the process compared to the LHC, and in this respect it is also very important for new physics studies, for example the anomalous trilinear gauge boson couplings. The FCC-hh is planned to reach an integrated luminosity of 30 ab$^{-1}$ at a center-of-mass energy of 100 TeV \cite{fcc}.

\begin{table}
\caption{The best limits at 95$\%$ Confidence Level (C.L.) on $c_{\tilde{W}WW}/\Lambda^2$, $c_{WWW}/\Lambda^2$, $c_{\tilde{W}}/\Lambda^2$, $c_{W}/\Lambda^2$ and $c_{B}/\Lambda^2$ parameters obtained from the LHC experiments.}
\label{tab1}
\begin{ruledtabular}
\begin{tabular}{cccccccc}
{Experimental limit} &$c_{\tilde{W}WW}/\Lambda^2$ & $c_{WWW}/\Lambda^2$  & $c_{W}/\Lambda^2$ & $c_{B}/\Lambda^2$ & $c_{\tilde{W}}/\Lambda^2$\\
& (TeV$^{-2}$) & (TeV$^{-2}$) & (TeV$^{-2}$)& (TeV$^{-2}$)& (TeV$^{-2}$)\\
\hline
ATLAS \cite{39} & [-5.30; 5.30] & $-$ & [-6.40; 11.0] & [-36.0; 43.0] & $-$\\
($WW/WZ\rightarrow l \nu_{l} q q'$) & & & & \\
($\sqrt{s}=8$ TeV and $L_{int}=20.2$ fb$^{-1}$) & & & & \\
CMS  \cite{40} & $-$ & [-5.70; 5.90] & [-11.4; 5.40]& [-29.2; 23.9]& $-$\\
($\sqrt{s}=8$ TeV and $L_{int}=19.4$ fb$^{-1}$) & & & & \\
CMS  \cite{cms} & $-$ & [-1.58; 1.59] & [-2.00; 2.65]& [-8.78; 8.54]& $-$\\
($\sqrt{s}=13$ TeV and $L_{int}=35.9$ fb$^{-1}$) & & & & \\
CMS  \cite{41} & [-0.90; 0.91] & [-0.45; 0.45] & [-39.7; 40.7]& [-39.7; 40.7]& [-20.3; 20.0]\\
($W\gamma \rightarrow l \nu_{l} \gamma$) & & & & \\
($\sqrt{s}=13$ TeV and $L_{int}=137$ fb$^{-1}$) & & & & \\
\end{tabular}
\end{ruledtabular}
\end{table}

The rest of our study is organized as follows: In Section II, we will present the production of the signal and the background events of the
process $pp \rightarrow W^{\pm} \gamma$ at the FCC-hh. The limits on the anomalous couplings will be obtained in Section III.
Finally, we conclude with the sensitivities of each anomalous coupling discussed in Section IV.

\section{Production of signal and background events}

The effects of the anomalous contributions arising from dimension-6 operators and SM contributions as well as interference between new physics and the SM contribution are performed via the process $pp \rightarrow W^{\pm} \gamma$ at the FCC-hh. For this purpose, the tree-level Feynman diagrams of the process $pp \rightarrow W^{\pm} \gamma $ are given in Fig.1. In this figure, the first diagram comprises the anomalous $WW\gamma$ couplings and the others come from the SM processes. In order to investigate the effects of the anomalous couplings on the process $pp \rightarrow W^{\pm} \gamma $, we use MadGraph5$-$aMC$@$NLO \cite{42} after implementation of the operators Eqs.(3)$-$(7) through Feynrules package \cite{43} as a Universal FeynRules Output (UFO) module
\cite{44}. The generated signal and relevant backgrounds in MadGraph5$-$aMC$@$NLO are passed through Pythia 8 for parton shower and hadronization \cite{45}. The detector responses are taken into account with Delphes package \cite{46}. We have used Delphes default parameters with the FCC-hh card for the detector simulation. However, all events are analyzed
using the ExRootAnalysis utility  \cite{47} with ROOT  \cite{48}.

\begin{figure}[t]
\centerline{
\scalebox{0.35}{\includegraphics{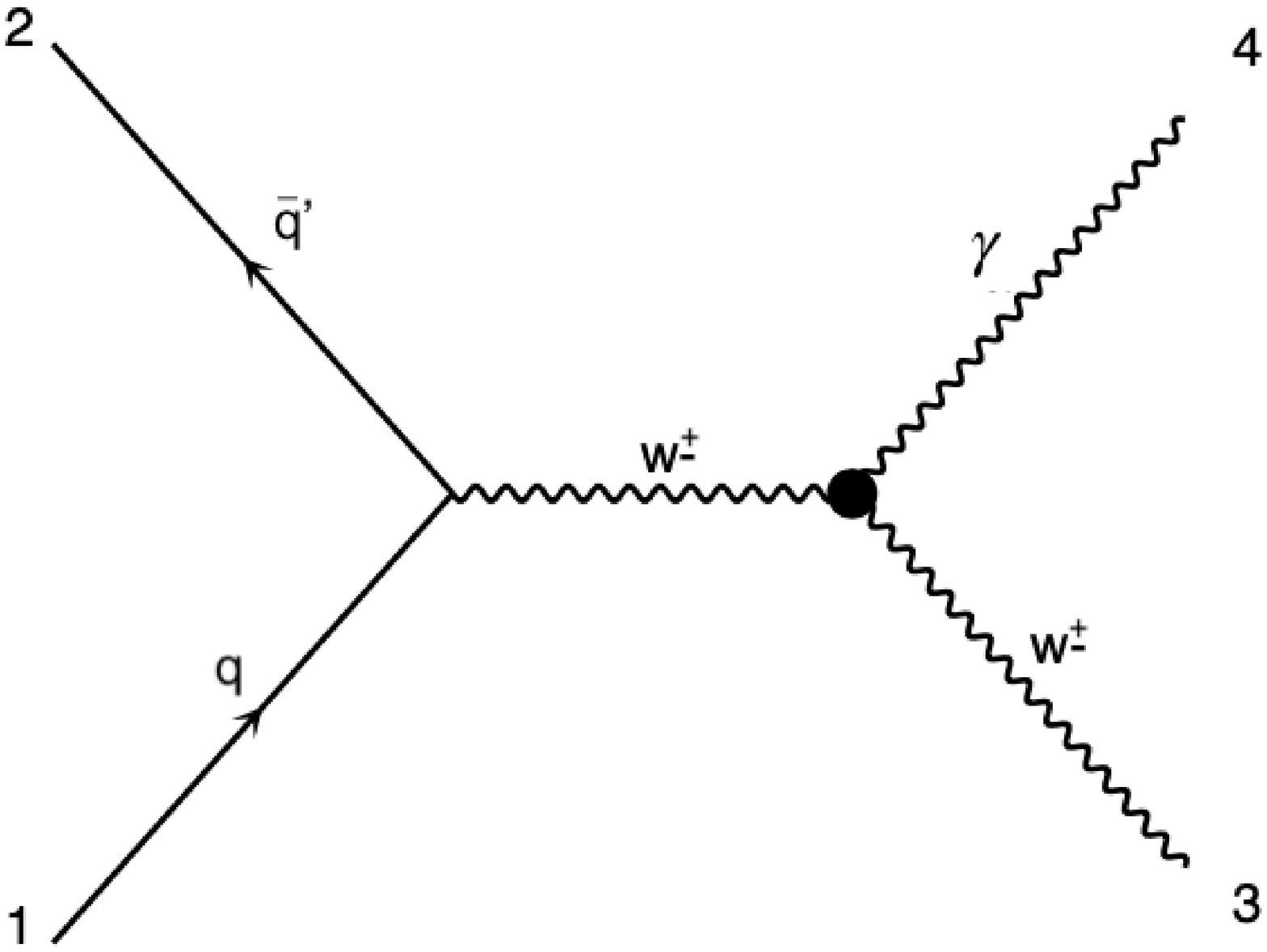}}
\qquad\; \scalebox{0.34}{\includegraphics{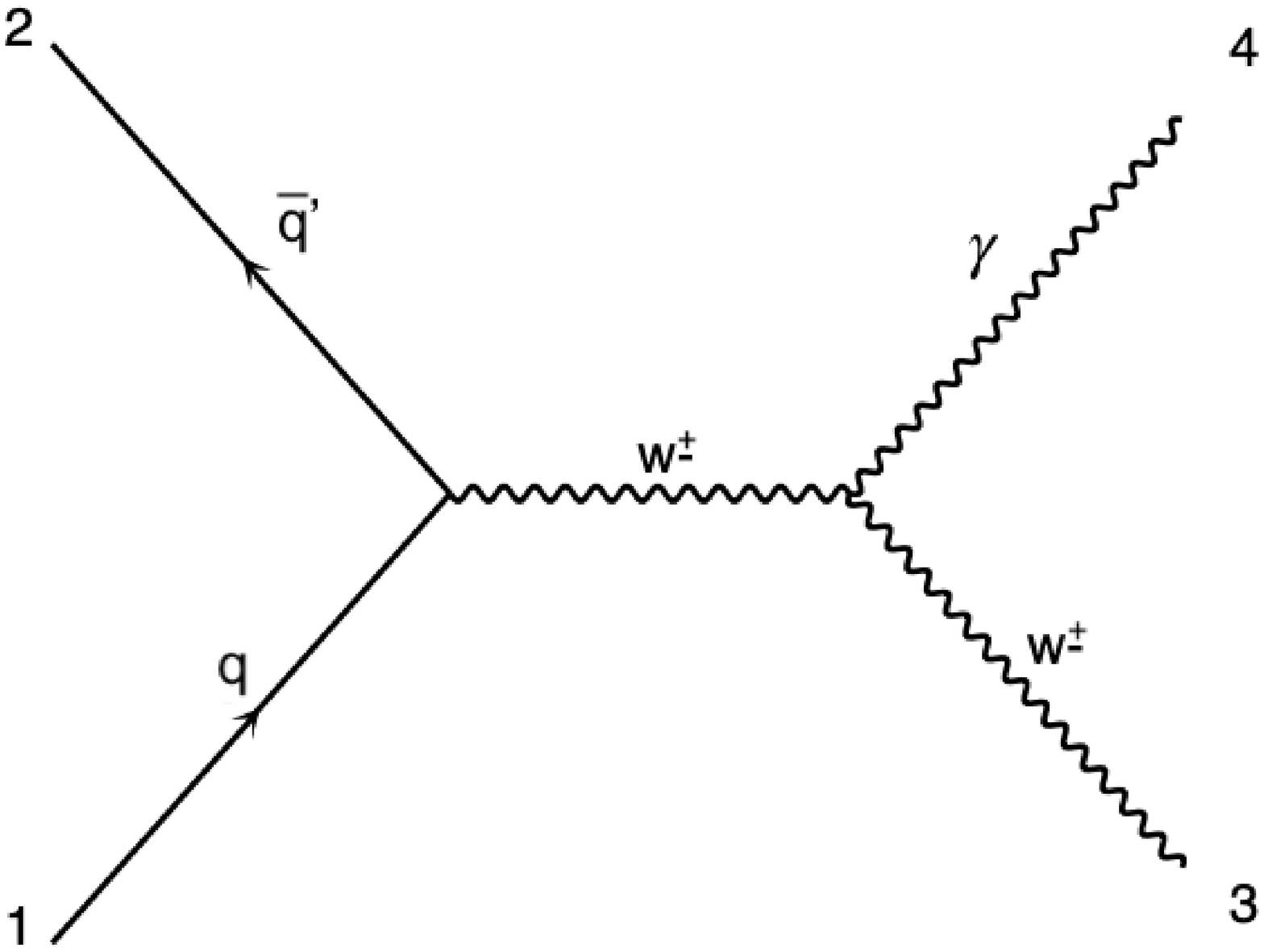}}}
\vspace{0.5cm}
\centerline{
\scalebox{0.34}{\includegraphics{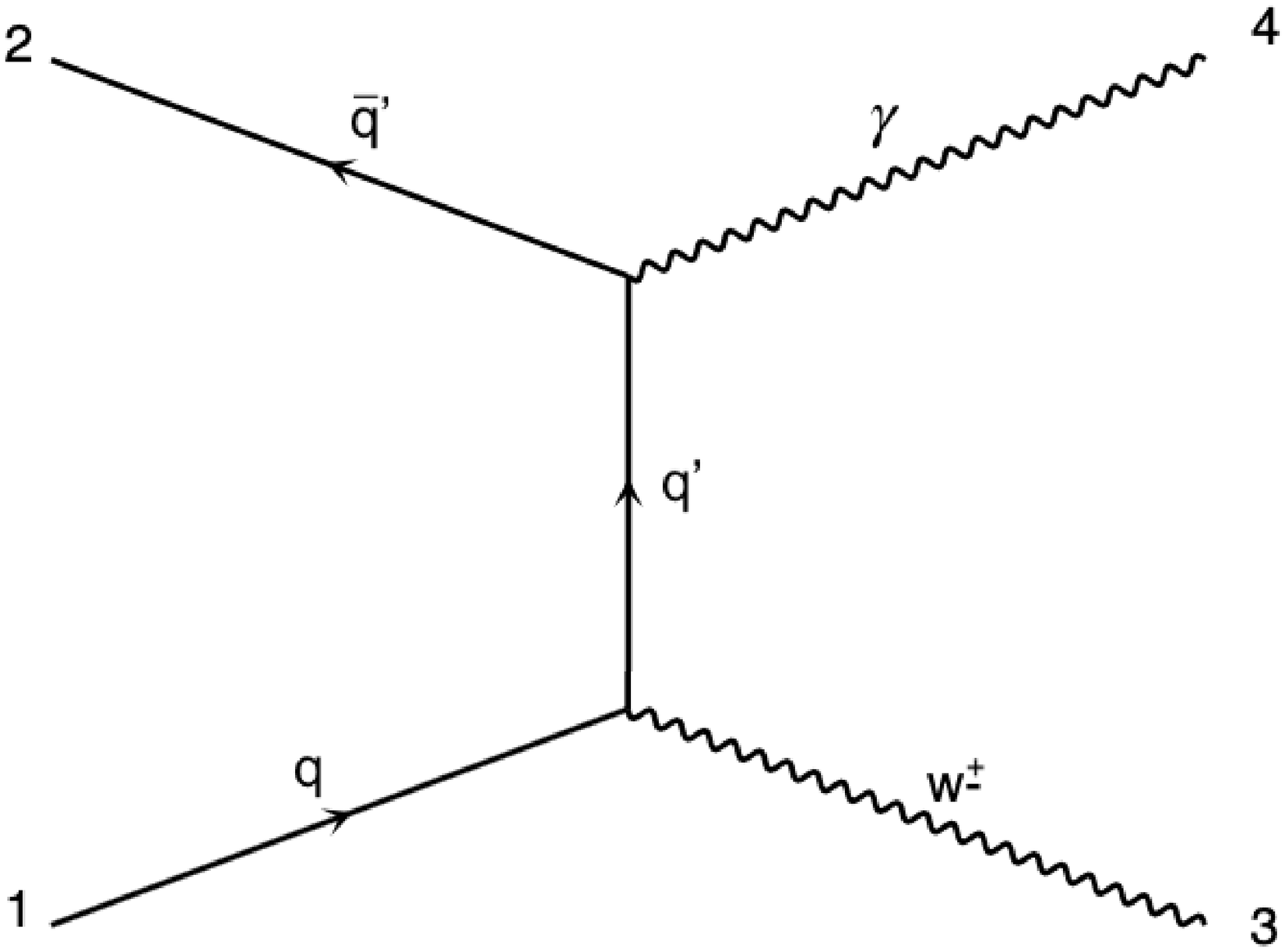}}
\qquad\; \scalebox{0.34}{\includegraphics{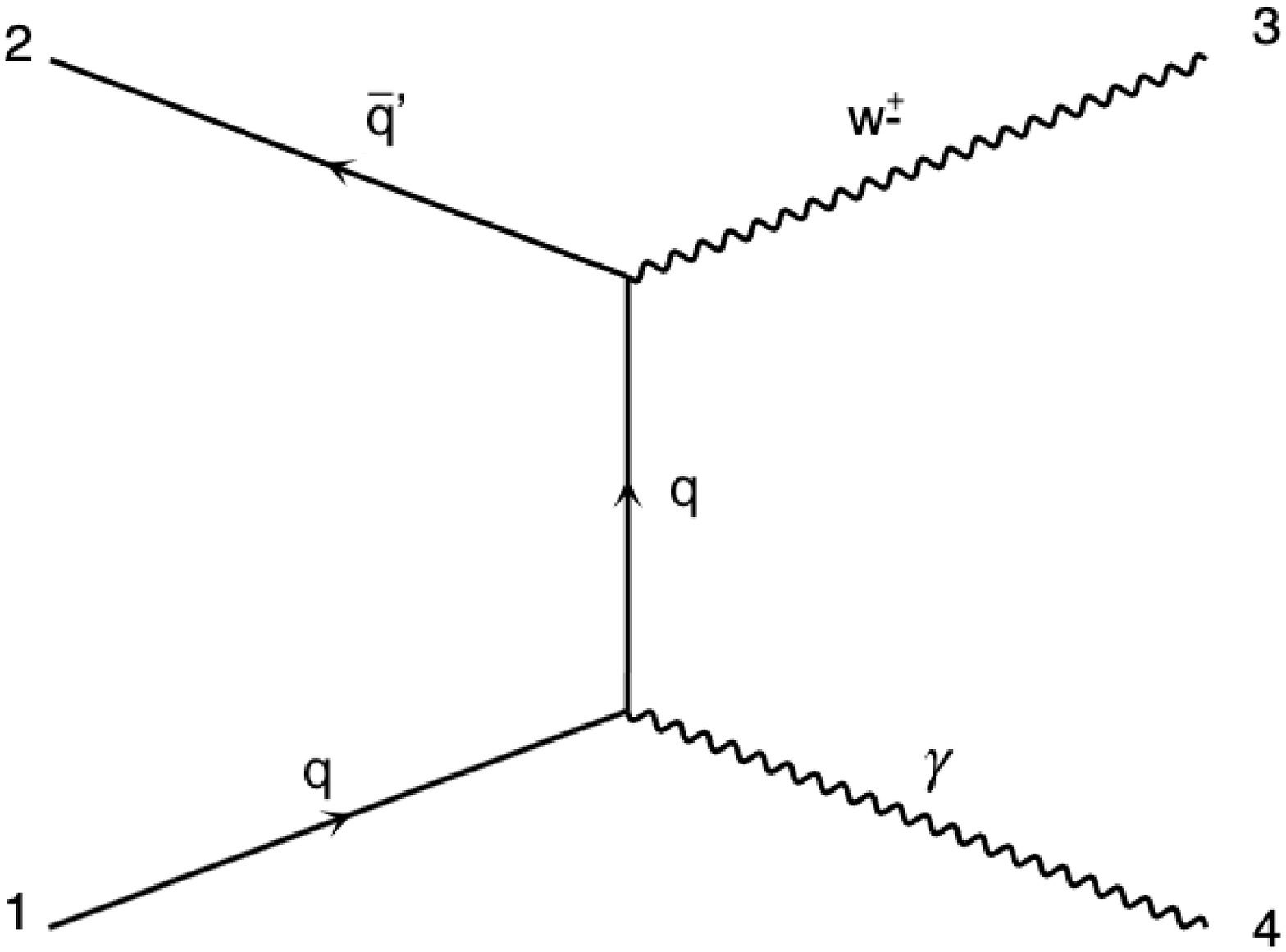}}}
\caption{ \label{1} Feynman diagrams for the subprocess
$q\bar{q}\to W^{\pm}\gamma$ including the effective $WW\gamma$ vertex.}
\end{figure}

In this work, we examine the anomalous $WW\gamma$ couplings through the process $pp \rightarrow W^{\pm} \gamma\rightarrow \ell^{\pm} \nu_{\ell} (\bar{\nu_{\ell}}) \gamma$ ($\ell^{\pm}= e^{\pm}, \mu^{\pm}$) at the FCC-hh. Here, the final state topology of the investigated process occurs a photon, a charged lepton and missing energy. Thus, as backgrounds, the following relevant SM background processes having the same or similar final state topology are taken into consideration;

\begin{eqnarray}
SM:pp\rightarrow \ell^{\pm} \nu_{\ell}(\bar{\nu_{\ell}}) \gamma \,,
\end{eqnarray}
\begin{eqnarray}
Z+\gamma:pp\rightarrow Z \gamma \,,
\end{eqnarray}
\begin{eqnarray}
W+j:pp\rightarrow W^{\pm} j \,,
\end{eqnarray}
\begin{eqnarray}
W+W+\gamma:pp\rightarrow W^{\pm} W^{\mp} \gamma \,,
\end{eqnarray}
\begin{eqnarray}
W+Z+\gamma:pp\rightarrow W^{\pm} Z \gamma \,,
\end{eqnarray}
\begin{eqnarray}
W+j+\gamma:pp\rightarrow W^{\pm} j \gamma \,,
\end{eqnarray}
\begin{eqnarray}
Z+j+\gamma:pp\rightarrow Z j \gamma \,,
\end{eqnarray}
\begin{eqnarray}
W+2j+\gamma:pp\rightarrow W^{\pm} j j \gamma \,,
\end{eqnarray}
\begin{eqnarray}
Z+Z+\gamma:pp\rightarrow Z Z \gamma \,.
\end{eqnarray}

We generate $W\gamma+jets$ and $Z\gamma+jets$ events with MG5 at the leading order. We apply a high  $p_T$ cut on the photon, then extra photons will be suppressed. For the preselection and kinematical cuts of leptons, photon and jets included in some backgrounds for the signal and the relevant backgrounds, we apply a minimum cuts (generator level):

\begin{itemize}
\item $p^l_T>10$ GeV, $p^j_T>20$ GeV, $p^\gamma_T>100$ GeV,
\item $|\eta^{l}| < 2.5$, $|\eta^{j}| < 5$, $|\eta^{\gamma}| < 2.5$,
\item $\Delta R( \gamma, \ell)>0.4$, $\Delta R( \gamma, j)>0.4$, $\Delta R( \ell, j) >0.4$.
\end{itemize}

Here, while $p_{T}^{l, j, \gamma}$ are the transverse momentums of the final state particles, $\eta^{l, j, \gamma}$ are the pseudorapidities of the final state particles. To have well-separated charged leptons, jets and the photon in the phase space, we require angular separations $(\Delta R =( (\Delta \phi)^2+ (\Delta \eta)^2)^{1/2}$). Actually, we do not require extra jets on the signal selection, however, in the background generation we allow at most one jet. We have not considered fake signal from the conversion jets to electrons or photons.

\begin{figure}[t]
\centerline{\scalebox{0.7}{\includegraphics{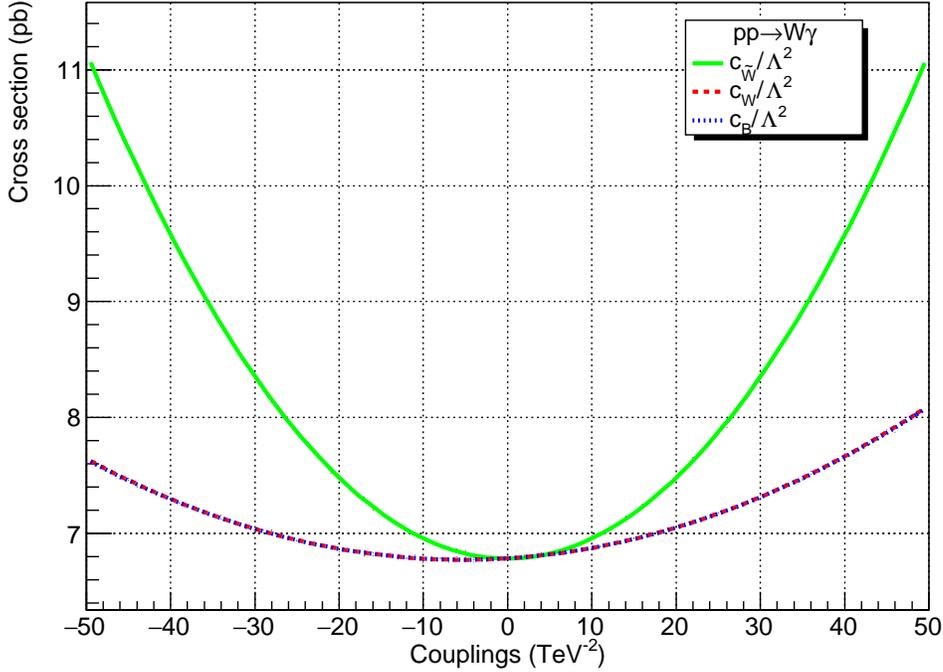}}}
\caption{ \label{2} The total cross-sections of the process $pp\to W^{\pm}\gamma$ as a function of the anomalous $C_{\tilde{W}}/ \Lambda^{2}$, $C_{W}/ \Lambda^{2}$ and $C_{B}/ \Lambda^{2}$ couplings at the FCC-hh.}
\end{figure}

\begin{figure}[t]
\centerline{\scalebox{0.7}{\includegraphics{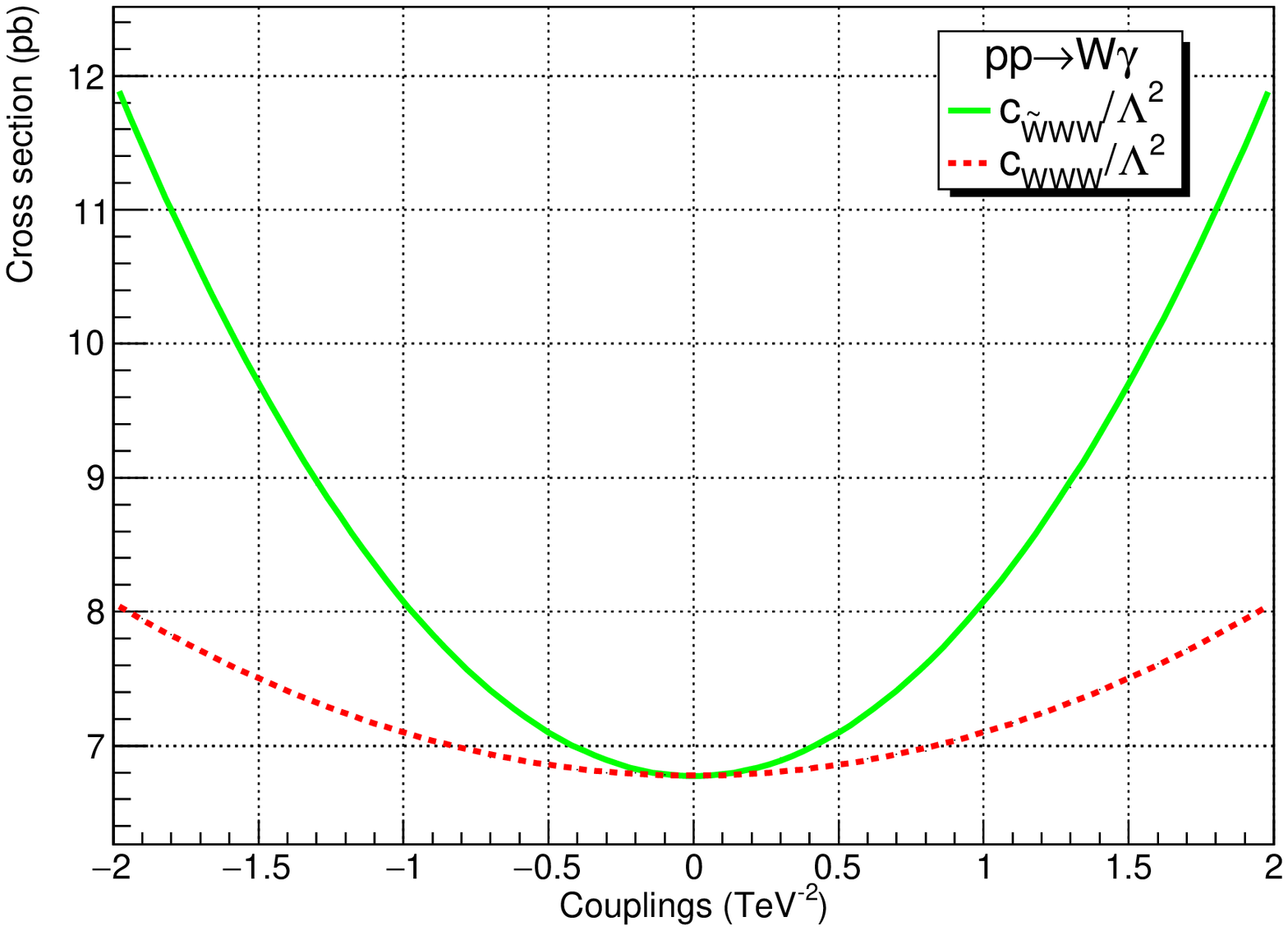}}}
\caption{ \label{2} The total cross-sections of the process $pp\to W^{\pm}\gamma $ as a function of the anomalous $C_{\tilde{W}WW}/ \Lambda^{2}$ and $C_{WWW}/ \Lambda^{2}$ couplings at the FCC-hh.}
\end{figure}

In Figs. 2 and 3, the total cross-sections of the process $pp \rightarrow W^{\pm}\gamma $ as a function of the anomalous $C_{\tilde{W}}/\Lambda^2$, $C_{W}/\Lambda^2$,$C_{B}/\Lambda^2$, $C_{\tilde{W}WW}/\Lambda^2$ and $C_{WWW}/\Lambda^2$ couplings are given. Here, one of the anomalous couplings is non-zero at any time, while the other couplings are fixed to zero.  As can be seen from Figs. 2 and 3, the total cross-sections of the process $pp \rightarrow W^{\pm}\gamma $ increase with increasing coupling values. For $C_{W}/\Lambda^2$ and $C_{B}/\Lambda^2$, we obtain degenerate behavior of the cross sections as seen in Fig. 2. The cross-sections after using the minimal cut set are given in Table II to show the effects of the signal and the relevant backgrounds for different values of the anomalous couplings. In this table, the cross-sections for the signal and the backgrounds in the relevant mode and decay channel are given. The final state topology of the signal is characterized as a high $p_{T}^{\gamma}$
associated with a lepton and missing transverse momentum from the undetected
neutrino. We also take into account the final states associated with a photon as background in the event
generation and account for the detector responses in the simulation.

\begin{table}[ht]
\caption{Cross sections for the signal and the backgrounds in the
relevant mode and decay channel at the FCC-hh.}
\begin{tabular}{|c|c|c|c|}
\hline
Couplings (TeV$^{-2}$) & Mode & Decay Channel & Cross Section (pb)\tabularnewline
\hline
\hline
$c_{WWW}/\varLambda^{2}=0.8$ & $W^{\pm}+\gamma$ & \multicolumn{1}{c|}{$l^{\pm}+\cancel{E_{T}}+\gamma$} & $9.6\times10^{-1}$\tabularnewline
\hline
$C_{W,B}/\Lambda^{2}=40$ & $W^{\pm}+\gamma$ & \multicolumn{1}{c|}{$l^{\pm}+\cancel{E_{T}}+\gamma$} & $1.1\times10^{0}$\tabularnewline
\hline
$c_{\tilde{W}WW}/\varLambda^{2}=0.4$ & $W^{\pm}+\gamma$ & \multicolumn{1}{c|}{$l^{\pm}+\cancel{E_{T}}+\gamma$} & $9.6\times10^{-1}$\tabularnewline
\hline
$c_{\tilde{W}}/\varLambda^{2}=20$ & $W^{\pm}+\gamma$ & \multicolumn{1}{c|}{$l^{\pm}+\cancel{E_{T}}+\gamma$} & $1.0\times10^{0}$\tabularnewline
\hline
\multirow{9}{*}{Backgrounds} & $W^{\pm}+\gamma$ & $l^{\pm}+\cancel{E_{T}}+\gamma$ & $9.2\times10^{-1}$\tabularnewline
\cline{2-4} \cline{3-4} \cline{4-4}
 & $W^{\pm}+j+\gamma$ & $l^{\pm}+\cancel{E_{T}}+j+\gamma$ & $7.8\times10^{0}$\tabularnewline
\cline{2-4} \cline{3-4} \cline{4-4}
 & $Z+\gamma$ & $2l+\gamma$ & $2.9\times10^{-1}$\tabularnewline
\cline{2-4} \cline{3-4} \cline{4-4}
 & $Z+j+\gamma$ & $2l+j+\gamma$ & $6.0\times10^{-1}$\tabularnewline
\cline{2-4} \cline{3-4} \cline{4-4}
 & $W^{+}+W^{-}+\gamma$ & $2l+\cancel{E_{T}}+\gamma$ & $5.3\times10^{-3}$\tabularnewline
\cline{2-4} \cline{3-4} \cline{4-4}
 & $Z+Z+\gamma$ & $2l+\cancel{E_{T}}+\gamma$ & $2.9\times10^{-4}$\tabularnewline
\cline{2-4} \cline{3-4} \cline{4-4}
 & $W^{\pm}+j$ & $l^{\pm}+\cancel{E_{T}}+j$ & $4.6\times10^{4}$\tabularnewline
\cline{2-4} \cline{3-4} \cline{4-4}
 & $W^{\pm}+2j+\gamma$ & $l^{\pm}+\cancel{E_{T}}+2j+\gamma$ & $1.2\times10^{1}$\tabularnewline
\hline
\end{tabular}
\end{table}

Thus, in order to distinguish signal from relevant backgrounds, we also apply the following analysis cuts:

\begin{itemize}
\item $N_{\ell} \geq 1$, $N_{\gamma} \geq 1$,
\item $p^l_T>35$ GeV, $|\eta^{l}| < 2.5$,
\item $p^\gamma_T>200$ GeV, $|\eta^{\gamma}| < 2.5$,
\item MET$> 40$ GeV.
\end{itemize}

Production of $W^{\pm}\gamma$ events with photons having large transverse momentum $p_{T}^{\gamma}$ is a sensitive search of new physics. The photon in the examined process has the advantage of being identifiable with high efficiency and purity. Also, as known, the high dimensional operators that define new physics parameters could affect the transverse momentum of the photon, particularly at high energy region of $p_{T}^{\gamma}$. This is very useful for distinguishing between signal and relevant backgrounds.  The numbers of expected events as a function of $p_{T}^{\gamma}$ photon transverse momentum for the $pp \to W^{\pm}\gamma $ signal and relevant backgrounds at the FCC-hh with $L_{int}=3$ ab$^{-1}$ and $\sqrt{s} = 100$ TeV are given in Fig. 4. We simulate signal ($pp \rightarrow W^{\pm} \gamma$) and interfering background ($pp \rightarrow W^{\pm} \gamma$) simultaneously, however other backgrounds are simulated separately. We have also used
other coupling values in the signal simulation and in the analysis to produce the plots for estimated sensitivities as shown in Fig. 5 and Fig. 6.

\begin{figure}[!h]
    \centering
    \includegraphics[width=8cm]{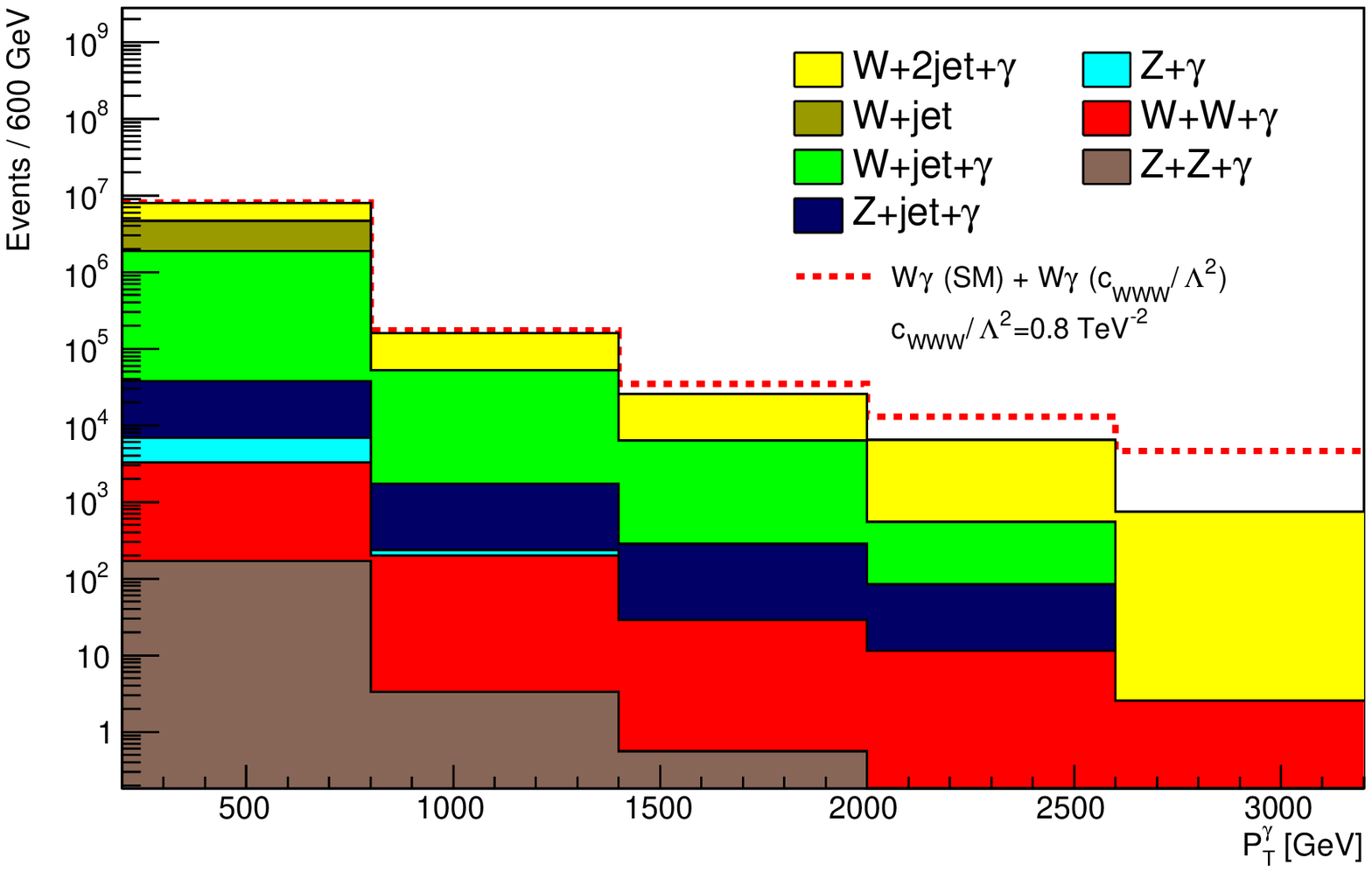} 
    \includegraphics[width=8cm]{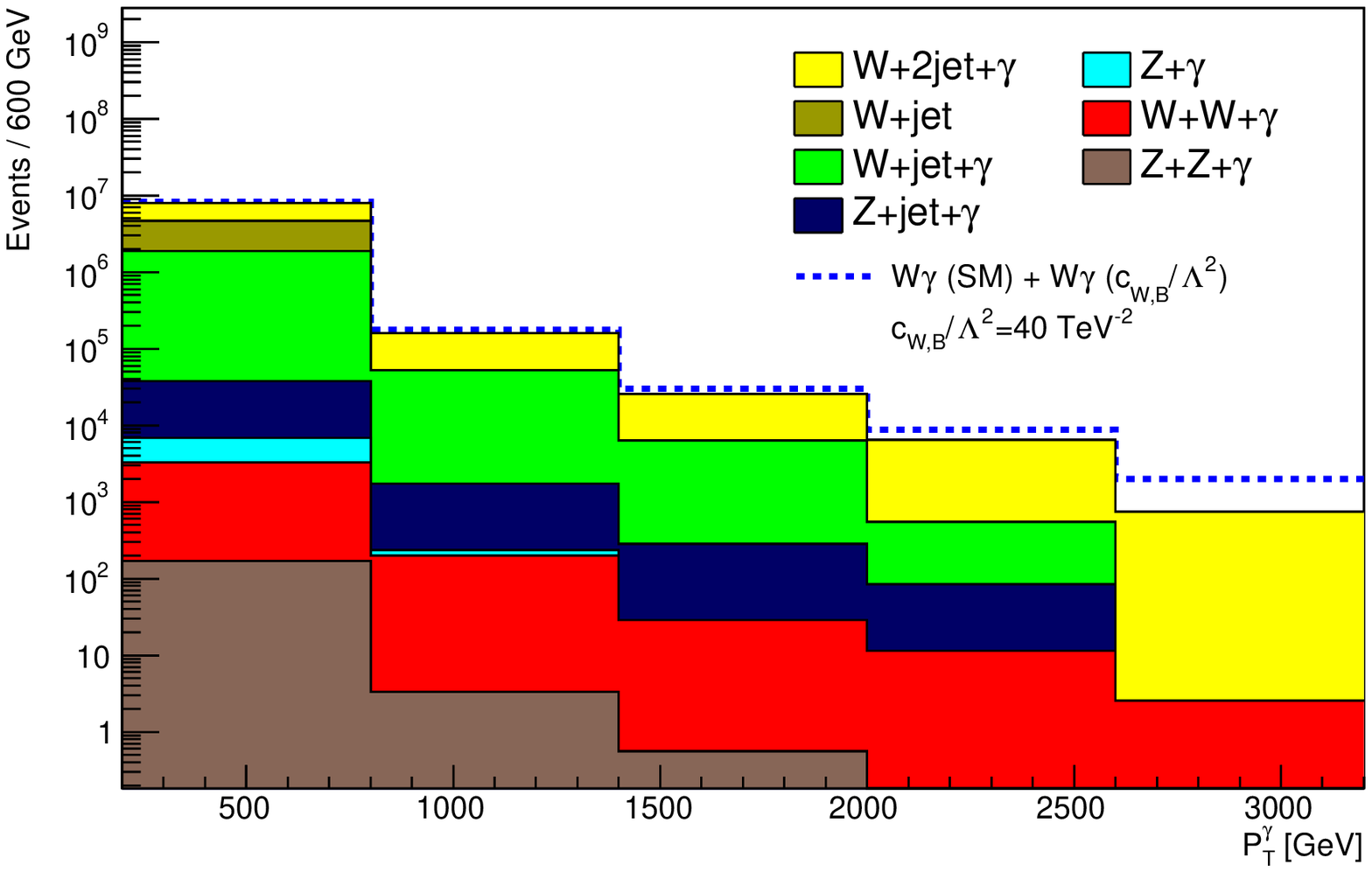}
    \includegraphics[width=8cm]{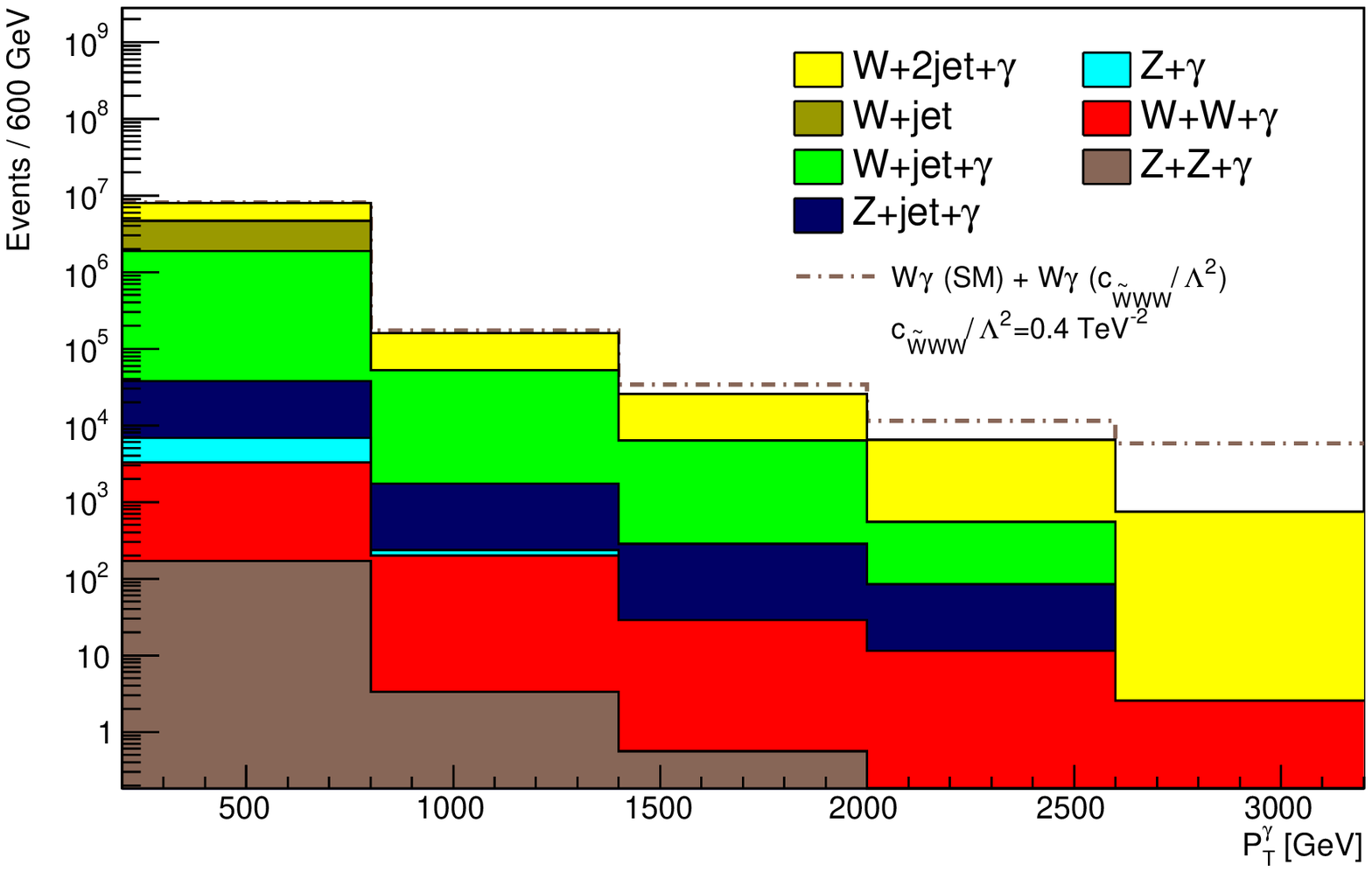} 
    \includegraphics[width=8cm]{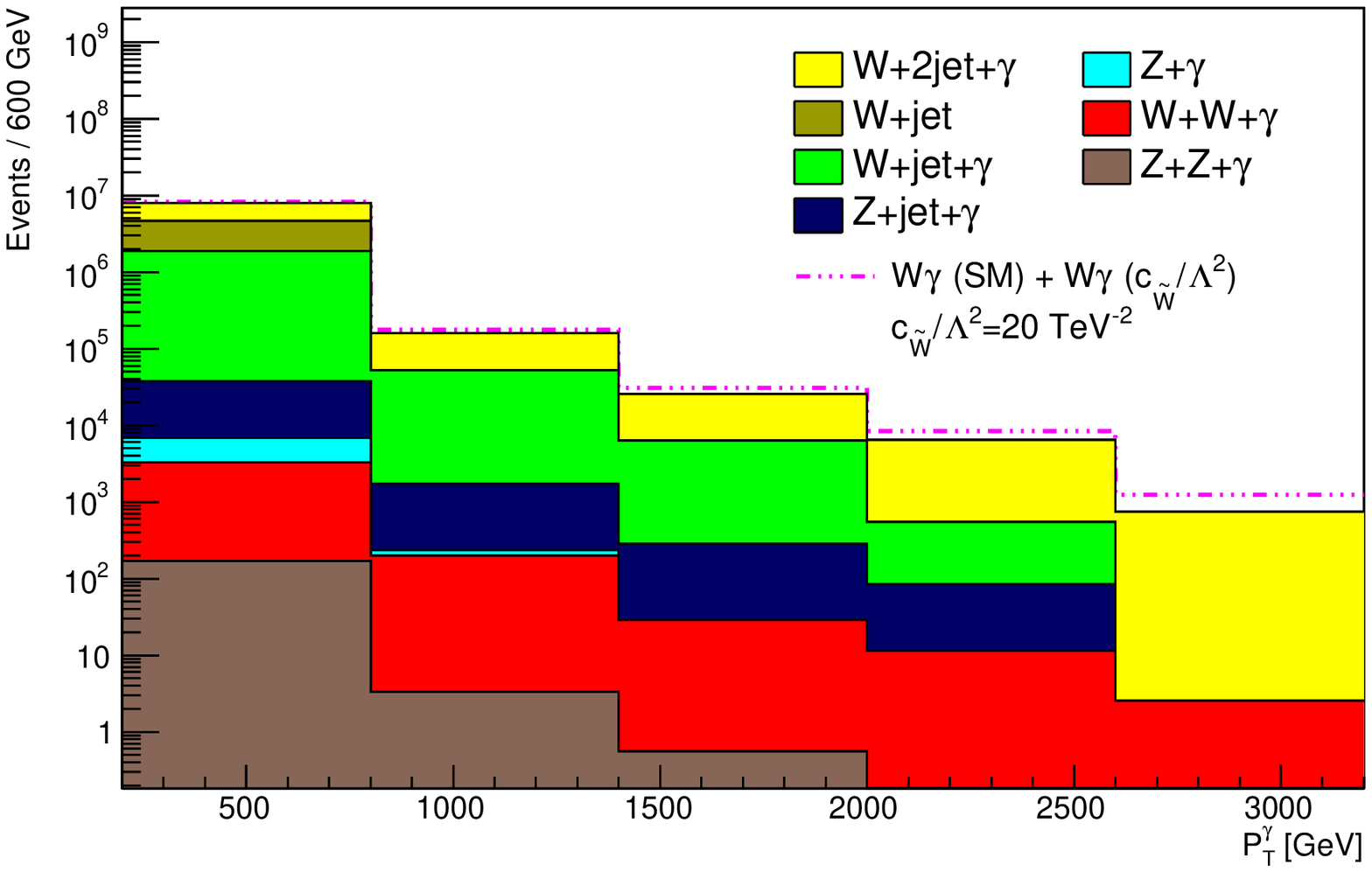} 
    \caption{ The transverse momentum of the photon $p_{T}^{\gamma}$  for the signals including certain values of the anomalous couplings and relevant backgrounds for the process $pp\to W^{\pm}\gamma$ at the FCC-hh. Here, the backgrounds are stack while the signal and interfering background are added (dashed lines).}
\end{figure}

\section{Evaluation of Results}
In order to obtain sensitivity to the anomalous $C_{\bar{W}}/\Lambda^2$, $C_{W,B}/\Lambda^2$, $C_{\tilde{W}WW}/\Lambda^2$ and $C_{WWW}/\Lambda^2$ couplings for the process $pp\to W^{\pm} \gamma $, we use a $\chi^{2}$ method, where the $\chi^{2}$ function with and without a systematic uncertainty is defined as follows

\begin{eqnarray}
\chi^{2} =\sum_i^{n_{bins}}\left(\frac{N_{i}^{NP}-N_{i}^{B}} {\Delta_i}\right)^{2}
\end{eqnarray}
where $N_i^{NP}$ is the total number of events in the existence of effective couplings, $N_i^B$ is total number of events of the
corresponding SM backgrounds in the $i$th bin of the $p_T^{\gamma}$ distributions, $\Delta_i=\sqrt{\delta_{sys}^2 {N_i^B}^{2}+N_i^B}$ includes both the systematic ($\delta_{sys}$) and statistical uncertainties in each bin.

The estimated sensitivities at $95\%$ C.L. on the anomalous $C_{\tilde{W}}/\Lambda^2$, $C_{W,B}/\Lambda^2$, $C_{\tilde{ W}WW}/\Lambda^2$ and $C_{WWW}/\Lambda^2$ couplings through the process $pp\to W^{\pm}\gamma$ at the FCC-hh with $\sqrt{s}=100$ TeV, $L_{int}=3$ ab$^{-1}$, $30$ ab$^{-1}$ and no-sys, $5\%$, $10\%$ are given in Tables III and IV. Also, in Figs.5 and 6, the achievable limits are shown as horizontal bar graphs. Our obtained 95\% C.L. limits without systematic uncertainties on the anomalous $C_{\tilde{W}}/\Lambda^2$, $C_{W,B}/\Lambda^2$, $C_{\tilde{W}WW}/\Lambda^2$ and $C_{WWW}/\Lambda^2$ couplings for the FCC-hh are $[-0.47; 0.47]$ TeV$^{-2}$, $[-0.88; 0.88]$ TeV$^{-2}$, $[-0.03; 0.03]$ TeV$^{-2}$ and $[-0.01; 0.01]$ TeV$^{-2}$, respectively.

\begin{table} [ht]
\caption{Attainable limits on $C_{W,B}/\Lambda^{2}$ and $C_{\tilde{W}}/\Lambda^{2}$ couplings from dimension-6 operator
 at $95\%$ C. L. for two different luminosity
projections of 3 ab$^{-1}$ and 30 ab$^{-1}$ at the FCC-hh.}
\begin{tabular}{|c|c|c|c|c|c|c|}
\hline
\multirow{2}{*}{Couplings (TeV$^{-2}$)} & \multicolumn{3}{c|}{$L_{int}=3$ ab$^{-1}$} & \multicolumn{3}{c|}{$L_{int}=30$  ab$^{-1}$}\tabularnewline
\cline{2-7} \cline{3-7} \cline{4-7} \cline{5-7} \cline{6-7} \cline{7-7}
 & no-sys & $5\%$ sys & $10\%$ sys & no-sys & $5\%$ sys & $10\%$ sys\tabularnewline
\hline
$C_{W,B}/\Lambda^{2}$ & [-2.93; 2.69] & [-21.00; 19.75] & [-33.74; 28.11] & [-0.88; 0.88] & [-20.95; 19.28] & [-33.74; 28.11]\tabularnewline
\hline
$C_{\tilde{W}}/\Lambda^{2}$ & [-1.46; 1.50] & [-10.88; 11.18] & [-15.53; 15.50] & [-0.47; 0.47] & [-10.77; 11.06] & [-15.53; 15.50]\tabularnewline
\hline
\end{tabular}
\end{table}

\begin{table} [ht]
\caption{Same as Table III but for $C_{WWW}/\Lambda^{2}$ and $C_{\tilde{W}WW}/\Lambda^{2}$ couplings.}

\begin{tabular}{|c|c|c|c|c|c|c|}
\hline
\multirow{2}{*}{Couplings (TeV$^{-2}$)} & \multicolumn{3}{c|}{$L_{int}=3$ ab$^{-1}$} & \multicolumn{3}{c|}{$L_{int}=30$ ab$^{-1}$}\tabularnewline
\cline{2-7} \cline{3-7} \cline{4-7} \cline{5-7} \cline{6-7} \cline{7-7}
 & no-sys & $5\%$ sys & $10\%$ sys & no-sys & $5\%$ sys & $10\%$ sys\tabularnewline
\hline
$C_{WWW}/\Lambda^{2}$ & [-0.03;0.03] & [-0.19;0.18] & [-0.35;0.34] & [-0.01;0.01] & [-0.18;0.17] & [-0.35;0.34]\tabularnewline
\hline
$C_{\tilde{W}WW}/\Lambda^{2}$ & [-0.05;0.04] & [-0.13;0.12] & [-0.17;0.16] & [-0.03;0.03] & [-0.13;0.12] & [-0.17;0.16]\tabularnewline
\hline
\end{tabular}
\end{table}

\begin{figure}[t]
\centerline{\scalebox{0.7}{\includegraphics{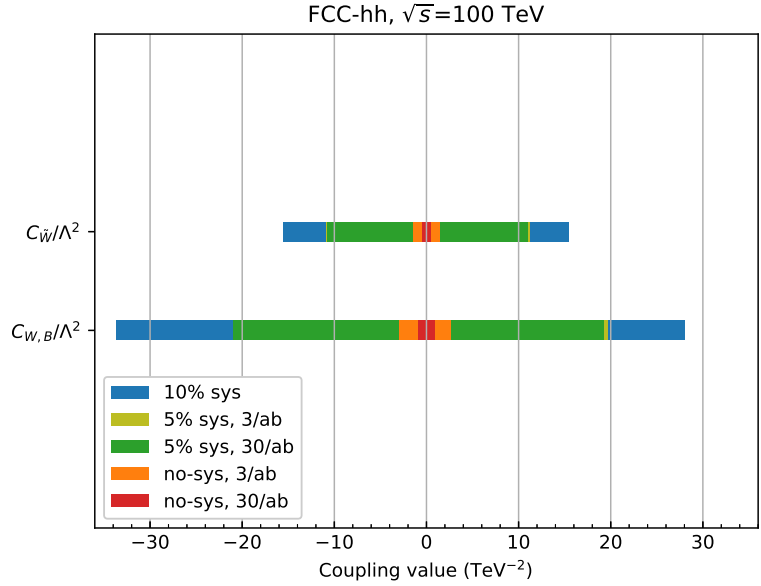}}}
\caption{The estimated sensitivities at $95\%$ C.L. on the anomalous $C_{\tilde{W}}/\Lambda^2$ and $C_{W,B}/\Lambda^2$ couplings through the process $pp\to W^{\pm}\gamma $ at the FCC-hh with $\sqrt{s}=100$ TeV, $L_{int}=3$ ab$^{-1}$, $30$ ab$^{-1}$ and no-sys, $5\%$ and $10\%$.}
\end{figure}

\begin{figure}[t]
\centerline{\scalebox{0.7}{\includegraphics{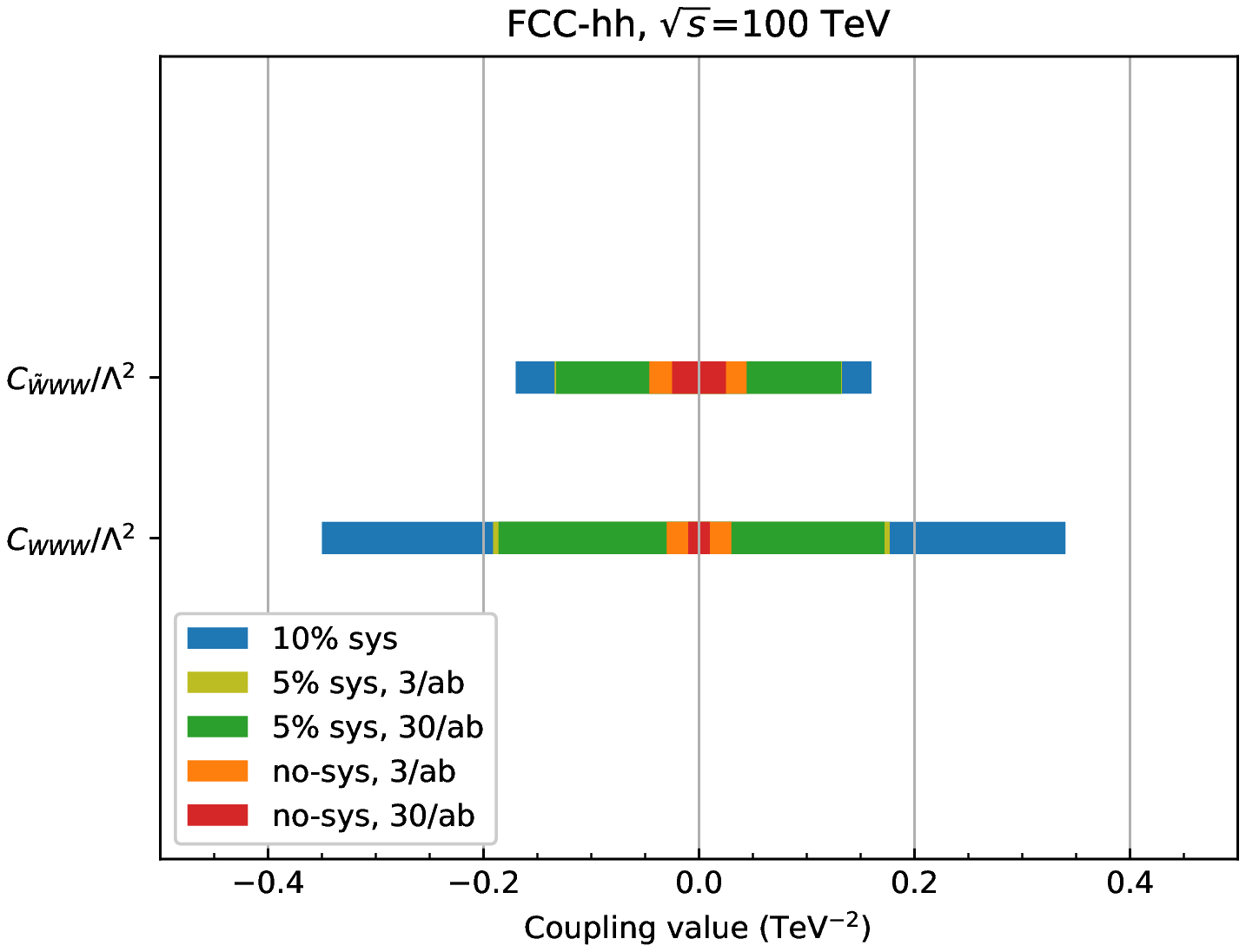}}}
\caption{Same as in Fig. 5 but for $C_{\tilde{W}WW}/\Lambda^2$ and $C_{WWW}/\Lambda^2$.}
\end{figure}

When we compare Table I, we see that we obtain better limits than current experimental limits. Here, $C_{WWW}/\Lambda^2$ coupling has the best sensitivity among the other couplings. For center-of-mass energy of 100 TeV with integrated luminosity 30 ab$^{-1}$, while we get the best limits on the order of O(10$^{-1}$) for $C_{\tilde{W}}/\Lambda^2$, $C_{W,B}/\Lambda^2$ we obtain the limits on the order of O(10$^{-2}$) for $C_{\tilde{ W}WW}/\Lambda^2$ and $C_{WWW}/\Lambda^2$ couplings in case of only statistical uncertainty. Also, our sensitivities obtained for $L_{int}=$30 ab$^{-1}$ without systematic uncertainty are almost up to 3 times better than derived $L_{int}=$3 ab$^{-1}$. For a systematic uncertainty of  $\delta_{sys}=10\%$, we estimate the sensitivities obtained on all couplings are better than the best experimental limit values. Also, our best limits on the anomalous couplings for $pp\to W^{\pm}\gamma $ with $\delta_{sys}=5\%$ can be improved a factor of 1.5-2 according to the case with $\delta_{sys}=10\%$. Figs.5 and 6 represent the limits on the anomalous couplings which do not increase proportionally to the increasing luminosity due to the systematic uncertainty
considered here. The reason of this situation is the systematic uncertainty which is much bigger than the statistical uncertainty. Thus, as can be seen Tables III-IV and Figs.5 and 6, if the systematic uncertainty is improved, we expect better limits on the anomalous couplings.

\section{Conclusions}

We have examined the effects of dimension-6 operators giving rise to the anomalous interactions of $WW\gamma$ vertex through the process $pp \rightarrow W^{\pm}\gamma $ at the FCC-hh. The advantage of the process $pp \rightarrow W^{\pm}\gamma $ is that it studies the $WW\gamma$ couplings independently of $WWZ$ effects. However, the high dimensional operators could affect $p_{T}$ distribution of the photon in the final state, especially at the region with large $p_{T}$ values, which can be very useful to distinguish signal and background events. Therefore, we use this as a tool to investigate the sensitivity to $C_{\tilde{W}}/\Lambda^2$, $C_{W,B}/\Lambda^2$, $C_{\tilde{W}WW}/\Lambda^2$ and $C_{WWW}/\Lambda^2$ couplings that are described by dimension-6 operators. Here, we suppose that only one anomalous couplings is non-zero at a time. We deduce that the FCC-hh collider will be able to provide sensitivity on the anomalous $WW\gamma$ couplings which can set more stringent limits by three (two) orders of magnitude with respect to the best sensitivities of $C_{\tilde{W}}/\Lambda^2$, $C_{W,B}/\Lambda^2$ ($C_{\tilde{W}WW}/\Lambda^2$ and $C_{WWW}/\Lambda^2$) couplings derived from CMS Collaboration for the process $pp\rightarrow W^{\pm}\gamma $ at the center-of-mass energy of 13 TeV and integrated luminosity of 137 fb$^{-1}$ \cite{41}. We also discuss the impact of systematic uncertainties on our results. Including the systematic uncertainties increase the size of the estimated uncertainty and thus decrease the sensitivity on the anomalous couplings. As can be seen from Tables III and IV, our best limits obtained at $\delta_{sys}=10\%$ systematic uncertainty are better than the current experimental limits \cite{41}. As a result, we emphasize that the sensitivities obtained on the anomalous $WW\gamma$ couplings in this work are better than the sensitivity of the present experimental limits.

\pagebreak

\newpage

\end{document}